\documentclass[11pt,preprintnumbers,aps,amssymb,nofootinbib,amsmath,superscriptaddress]
{revtex4}
\usepackage{epsfig,epsf}
\usepackage{bm} 
\usepackage{epstopdf}
\newcommand{\beq}{\begin{equation}}
\newcommand{\beql}[1]{\begin{equation}\label{#1}}
\newcommand{\eeq}{\end{equation}}
\newcommand{\eq}[1]{(\ref{#1})}
\newcommand{\fig}[1]{Fig.~\ref{#1}}
\renewcommand{\sec}[1]{Sec.~\ref{#1}}
\newcounter{topiccounter}
\renewcommand{\b}[1]{{\bm #1}} 
\newcommand{\unit}[1]{\hat {{\bm #1}}} 

\newcommand{\e}{\varepsilon}

\begin{document}



\title{Magnetic contribution to dilepton production in heavy-ion collisions }

\author{Kirill Tuchin}

\affiliation{
Department of Physics and Astronomy, Iowa State University, Ames, IA 50011}

\date{\today}

\pacs{}

\begin{abstract}

We calculate a novel ``magnetic contribution" to the dilepton spectrum in heavy-ion collisions arising from interaction of relativistic quarks with intense magnetic field.  
Synchrotron radiation by quarks, which can be approximated by the equivalent photon flux, is followed by dilepton decay of photons in intense magnetic field. We argue that  ``magnetic contribution" dominates the dilepton spectrum at low lepton energies, whereas a conventional photon dilepton decay  dominates at higher lepton energies. 

\end{abstract}

\maketitle

\section{Introduction}\label{sec:int}

Electromagnetic radiation that accompanies any relativistic heavy ion collision weakly interacts with hot nuclear matter. However, it strongly interacts with 
highly intense magnetic field that is frozen into the nuclear matter \cite{Itakura:2013cia,Tuchin:2012mf,Tuchin:2010gx,Yee:2013qma}. Therefore it bears witness to 
the magnetic field existence and provides a rare opportunity to experimentally study its properties. In practice, electromagnetic radiation caused by magnetic field is masked by electromagnetic radiation of quark-gluon plasma (QGP). It is not an easy problem to disentangle the two contributions. A good theoretical control over both is required to achieve this goal. 

In the present article we address the problem of dilepton production in heavy-ion collisions due to magnetic field.  Our work is partly motivated by  the resent experimental results  on dileptons and photons produced in heavy-ion collisions, which challenge the prevailing theories  of electromagnetic processes in high energy nuclear physics that neglect strong electromagnetic interactions. It was observed in \cite{Adare:2008ab,Adare:2009qk} that there is a significant underestimation of electromagnetic spectra by  theoretical models in the low momentum region. This indicates that there are additional contributions that have not been taken into account. We dub the additional lepton production due to magnetic field as the ``magnetic contribution" as opposed to the ``conventional contributions".

\begin{figure}[ht]
      \includegraphics[height=2.2cm]{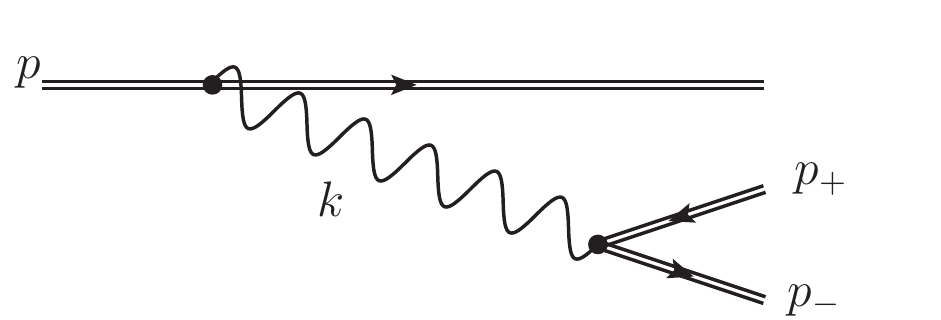} 
  \caption{Dilepton production by quark  in external magnetic field, mediated by a virtual photon.  Double lines denote fermion propagator in magnetic field. Components of four-vectors: $p=\{ \e,\b p\}$, $k=\{ \omega, \b k\}$, and $p_\pm=\{\e_\pm,\b p_\pm\}$.}
\label{fig:int1}
\end{figure}

\begin{figure}[ht]
      \includegraphics[height=2.2cm]{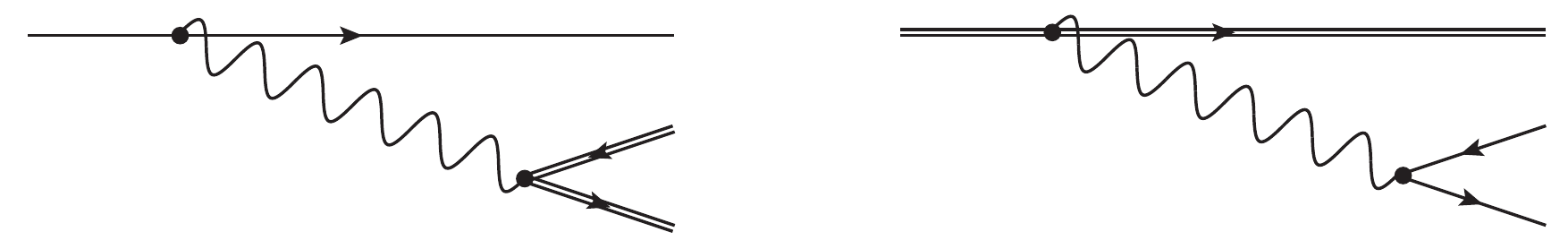} 
  \caption{Magnetic contributions that are not considered in this article.}
\label{fig:int2}
\end{figure}

Dilepton production in magnetic field proceeds in two stages. Firstly, quark (or antiquark) is produced in a heavy-ion collision. We will refer to the quark (or antiquark) distribution at this stage as the \emph{initial quark distribution}. 
Secondly, it radiates the lepton--anti-lepton pair in a processes mediated by a virtual photon as depicted in \fig{fig:int1} and \fig{fig:int2}.  Double lines  indicate quark and lepton propagators in magnetic field. 
In \fig{fig:int1} both quark and lepton move in magnetic field, whereas in \fig{fig:int2} only one of them does. The final result is sum over all these processes. However, in this article we will focus on the square of the amplitude shown in \fig{fig:int1}, because we anticipate that it is most sensitive to magnetic field. Other contributions will be considered elsewhere. 

The initial quark distribution is made up of soft and hard quarks and antiquarks. Soft  quarks have typical energies  $\e\sim T$, where $T$ is the plasma temperature, and are a part of QGP. The exact mechanism of their production and equilibration is not fully understood and is actually not essential  for our arguments.  What is important is an observation that according to the state-of-the-art phenomenology, the QGP  equilibration happens over  very short time on the order of $1/Q_s$, where $Q_s$ is the saturation momentum, see e.g.\ \cite{Kolb:2003dz}. Hard quarks in the central rapidity region $y=0$ have typical energies $\e \sim Q_s\gg T$  and their production mechanism has been thoroughly investigated \cite{Kharzeev:2008cv,Tuchin:2004rb,Kovchegov:2006qn,Gelis:2003vh,Blaizot:2004wv,Kopeliovich:2002yv,Kharzeev:2003sk}. In this article we focus on lepton energies $\e_+$ in the interval  $T< \e_+<Q_s$. In this region,  spectrum of soft quarks falls off exponentially, whereas the spectrum of hard quarks is only logarithmic. Moreover, the number  of soft quarks in plasma of volume $\mathcal{V}$ is on the order of $T^3 \mathcal{V}$. At early times $\mathcal{V}\sim \mathcal{S}/Q_s$ where $\mathcal{S}$ is the cross-sectional area of the ion overlap region. On the other hand, the number of hard quarks is on the order of $\mathcal{S}Q_s^2$ (see \sec{sec:aa}), which is much larger than the number of soft quarks. Therefore,  the contribution of hard quarks to dilepton production is dominant.  It is this contribution that we discuss in this article.

The paper is organized as follows. In \sec{sec:ww} we employ the Weisz\"acker-Williams method to write the dilepton production rate by a hard quark as a convolution of the  real photon decay rate with the flux of equivalent photons emitted by a fast quark. In \sec{sec:dec} we compute the rate of photon dissociation into dilepton as a function of lepton transverse momentum and rapidity. Initial quark distribution produced in heavy-ion collisions is computed in \sec{sec:aa} using the quasi-classical approximation. In \sec{sec:xx}  we apply the developed formalism to calculate the magnetic contribution to $e^+e^-$ pair production at midrapidity at RHIC. We compare dilepton production in magnetic field and in vacuum in \sec{sec:n} and argue that magnetic field contribution dominates at photon energies $\omega<200$~MeV. Finally, we summarize in \sec{sec:sum}.

\section{Equivalent photon approximation}\label{sec:ww}

Calculation of dilepton spectrum significantly simplifies because light quarks are ultra-relativistic $\e\gg m_q$ in the center-of-mass frame. This allows us to employ  the Weisz\"acker-Williams method to calculate the dilepton production with logarithmic accuracy. According to this method, we can relate the cross section of $q\to q\ell^+\ell^-$ process to the cross section of photo-production  $\gamma\to  \ell^+\ell^-$. The logarithmically enhanced contribution arises from the kinematic region where  photon virtuality has negligible effect both on photon emission and on dilepton photo-production.  The dilepton production rate can be written as (see notations in \fig{fig:int1})
\begin{align}\label{c1}
\frac{dN_{q\to \ell^+\ell^-}}{dt\,d\Omega_+ d\e_+}= \int  n(\omega)\, \frac{dN_{\gamma\to \ell^+\ell^-}}{dtd\Omega_+ d\e_+}\,d\omega \,,
\end{align}
where $n(\omega)$ is the flux of equivalent real photons replacing the virtual photon. It is  given by \cite{Baier:1973} 
\begin{align}\label{c3}
n(\omega)= \frac{2z_q^2\alpha}{\pi}\frac{1}{\omega}\ln\frac{\e}{\omega[1+(\chi\e/\omega)^{1/3}]}\,,
\end{align}
where $z_q$ is the quark's charge and
\begin{align}\label{c5}
\chi =\frac{e z_q}{m_q^3}\sqrt{-(F_{\mu\nu}p^\nu)^2} = \frac{e z_q}{m_q^3}\,|\b p\times \b B| = \frac{e z_q}{m_q^3} p_t  B\,
\end{align}
is a boost-invariant parameter. Here $\b p_t$ is quark's momentum component transverse to $\b B$  (not to be confused with $\b p_\bot$, which is transverse to the heavy-ion collision axis). Eq.~\eq{c1} takes into account  that the spectrum of photons is enhanced at $\omega\ll \e$. Angular distribution of photons strongly peaks at small angles $\sim m_q^2/\e^2$. Therefore, the equivalent photon momentum $\b k$ is approximately collinear with the quark's momentum $\b p$.  

It is convenient to introduce dimensionless parameters $b= \alpha^{1/2}B/m_q^2$, which is the value of magnetic field in units of the Schwinger's critical field, and $\gamma= \e/m_q$. In terms of these parameters $\chi= b\gamma \sin\theta$, where $\theta$ is the angle between $\b p$ and $\b B$. In heavy-ion collisions $\gamma\gg 1$ and $b\gtrsim 1$  implying that $\chi\gg 1$. This estimate breaks down only at very small angles $\theta< (b\gamma)^{-1}$ where $\chi$ becomes small and the photon decay rate is exponentially suppressed \cite{Berestetsky:1982aq} and hence can be neglected.
Integration in \eq{c1} runes over frequencies $2m_\ell\le \omega\le \omega_m$
with $\omega_m$ satisfying the equation
\beql{c9}
\omega_m \left[1+\left(\chi\e/\omega_m\right)^{1/3}\right]=\e\,.
\eeq
Since $\omega\ll \e$ and $\chi\gg 1$   it follows that $\e\chi/\omega\gg 1$. Using this in \eq{c9} we get 
\beql{c10}
\omega_m\approx  m_q \sqrt{\frac{\gamma}{b}}\,.
\eeq

To determine the region of applicability of the equivalent photon approximation, note that the logarithmic contribution to the equivalent photon  flux comes about only if  $\omega_m\gg m_\ell$. In view of \eq{c10} we have
\beql{c11}
\frac{\gamma}{b}\gg \frac{m_\ell^2}{m_q^2}\,.
\eeq
This condition is satisfied for not too strong fields and for light leptons. In particular, we will apply the equivalent photon approximation to calculate the magnetic contribution to electron-positron pair production at RHIC in \sec{sec:xx}.

\section{Photon dissociation rate}\label{sec:dec}

Now we turn to the photon dissociation rate which is convenient to perform in a frame where photon's momentum is perpendicular to the magnetic field. We denote such frame 
as $K'$ and all quantities in it bear the prime.  The corresponding dilepton rate is given in \sec{sec:dec-a}. In \sec{sec:dec-b} the rate is transformed to frame $K$ where photon moves at an arbitrary angle with respect to magnetic field. 

\subsection{$K'$-frame: $\b k'\cdot \b B'=0$. }\label{sec:dec-a}

In reference frame $K'$ photon moves in a plane perpendicular to the magnetic field:  $\b k'\cdot \b B'=0$. 
The rate of dilepton photo-production  in $K'$ frame reads \cite{Berestetsky:1982aq}
\begin{align}\label{c15}
\frac{dN_{\gamma\to \ell^+\ell^-}}{dt'd^3p'_+}=\frac{\alpha}{(2\pi)^2\omega'}\int_{-\infty}^\infty
d\tau &\left[ \frac{m_\ell^2}{\e_+'\e_-'}-(\e_+'^2+\e_-'^2)\omega_0^2\tau^2\frac{1}{4\e_-'^2}\right]\nonumber\\
&\times\exp\left\{ i\frac{\e_+'}{\e_-'}\omega'\tau\left( 1-\unit k'\cdot \b v_+'+\frac{\omega_0^2\tau^2}{24}\right)\right\}\,,
\end{align}
where  $\e_+'+\e_-'=\omega'$ and the synchrotron frequency is
\beql{c19}
\omega_0= \frac{eB'}{\e_+'}\,.
\eeq
 Integration over the time parameter $\tau$ can be done explicitly using the following formulas
\begin{align}
&\int_{-\infty}^\infty  \cos[b\tau +a\tau^3]d\tau= \frac{2\pi}{(3a)^{1/3}}\,\text{Ai}\!\left( \frac{b}{(3a)^{1/3}}\right)\,,\label{c21}\\
&\int_{-\infty}^\infty  \tau^2 \cos[b\tau +a\tau^3]d\tau=-\frac{b}{3a}\int_{-\infty}^\infty \cos[b\tau +a\tau^3]d\tau \,,\label{c23}
\end{align}
where $\text{Ai}(x)$ is the Airy function. Using \eq{c21},\eq{c23} into \eq{c15} we obtain
\begin{align}\label{c25}
\frac{dN_{\gamma\to \ell^+\ell^-}}{dt'd^3p_+'}=\frac{\alpha}{(2\pi)^2\omega'}&\left[ \frac{m_\ell^2}{\e_+'\e_-'}+2\frac{\e_+'^2+\e_-'^2}{\e_-'^2}(1-\unit k'\cdot \b v_+')\right]
4\pi \left( \frac{\e_-'}{\e_+'\omega' \omega_0^2} \right)^{1/3}
\nonumber\\
&\times\, \text{Ai}\!\left(2(1-\unit k'\cdot \b v_+')\left(\frac{\e_+'\omega'}{\e_-'\omega_0}\right)^{2/3}\right)\,.
\end{align}
Integration over the lepton direction, given by velocity $\b v_+'$, is convenient to do  in \eq{c15} before taking the $\tau$-integral. It can be performed with the required accuracy along the steps outlined  in \cite{Berestetsky:1982aq}. The main contribution arises from radiation of collinear lepton, i.e.\ $\unit k'\cdot \b v_+'\approx 1$. The result is 
\beql{c26}
\frac{dN_{\gamma\to \ell^+\ell^-}}{dt'd\e_+'}= \frac{\alpha m_\ell^2}{\omega'^2}\left\{ \int_{x'}^\infty \text{Ai}(\xi)d\xi + \left( \frac{2}{x'}-\varkappa x'^{1/2}\right)\text{Ai}'(x')\right\}\,,
\eeq
where $\text{Ai}'(x)$ is the derivative of the Airy function and
\beql{c26.1}
x'= \left( \frac{m_\ell^3\omega'}{eB'\e_+'\e_-'}\right)^{2/3}\,,\quad \varkappa = \frac{eB'\omega'}{m_\ell^3}\,.
\eeq
Parameter $\varkappa $ is boost-invariant. In an arbitrary frame   it reads  as follows  
\beql{c26.2}
\varkappa =\frac{e}{m_\ell^3}\sqrt{-(F_{\mu\nu}k^\nu)^2} = \frac{e}{m_\ell^3}\,|\b k\times \b B|\,.
\eeq
The total rate is
\beql{c26.3}
\frac{dN_{\gamma\to \ell^+\ell^-}}{dt'}=-\frac{\alpha\, eB'}{m_\ell\varkappa }\int_{(4/\varkappa )^{2/3}}^\infty\frac{2(x'^{3/2}+1/\varkappa )\,\text{Ai}'(x')}{x'^{11/4}(x'^{3/2}-4/\varkappa )^{1/2}}dx'\,.
\eeq

\subsection{$K$-frame: $\b k\cdot \b B\neq 0$. }\label{sec:dec-b}


Now we need to transform equations \eq{c1},\eq{c3},\eq{c15} to an arbitrary frame $K$. 
This is done by making a boost in the magnetic field direction. Let $z$ be the collision axis and $y$ be the magnetic field direction. We will use the following notations:  $\alpha$ and $\beta$ are polar and azimuthal angles of photon and quark with respect to  $\unit z$ axis (collision axis); $\theta$ and $\phi$ are polar and azimuthal angles of photon and quark with respect to $\unit y$ axis (magnetic field direction). The same symbols with the ``$+$" subscript refer to lepton $\ell^+$.
For example, quark  momentum reads
\begin{align}
\b p&= \e(\unit x \sin\alpha\cos\beta+\unit y\sin\alpha\sin\beta+\unit z\cos\alpha)\label{x1}\\
&= \e(\unit x \sin\theta\cos\phi+\unit y\cos\theta+\unit z\sin\theta\sin\phi)\,,
\label{x2}
\end{align}
and similarly for other vectors. 

 Suppose that $K$ moves with velocity $\b V= V\unit y$ with respect to $K'$. Boost along the magnetic field direction does not change the field, i.e $\b B'=\b B$. The Lorentz transformation formulas for the quark momentum read
\beql{c27}
p_x'=p_x\,,\quad p_y'= \gamma_V(p_y+V\e)\,,\quad p_z'=p_z\,,\quad \e'= \gamma_V(\e+Vp_y)\, 
\eeq
Since momenta of photon and quark are approximately collinear, in $K'$-frame $p'_y\approx k_y'=0$. This implies using \eq{c27},\eq{x2} that
\beql{c29}
V= -\cos\theta= -\sin\alpha\sin\beta\,,   \qquad   \gamma_V=(1-V^2)^{-1/2}=\frac{1}{\sin\theta}  \,.
\eeq
Transformation of quark and photon energies is given by
\beql{c31}
\e'= \e\sin\theta= \frac{\e}{\gamma_V}\,,\qquad \omega'=\frac{\omega}{\gamma_V}\,,
\eeq
Analogously to \eq{c27} we can write the transformation of the positively charged lepton's $\ell^+$ momentum 
\beql{c32}
p'_{+x}=p_{+x}\,,\quad p'_{+y}= \gamma_V(p_{+y}+V\e_+)\,,\quad p'_{+z}=p_{+z}\,,\quad \e'_+= \gamma_V(\e_++Vp_{+y})\,. 
\eeq
Thus, lepton's energy $\e_+$  transforms as 
\beql{c33}
\e'_+= \gamma_V\e_+(1+V\cos\theta_+)= \gamma_V\e_+(1+V \sin\alpha_+\sin\beta_+)\,.
\eeq
Consider a relativistic invariant 
\beql{c35}
k\cdot p_+= \omega\e_+(1-\unit k\cdot \b v_+)= \omega'\e'_+(1-\unit k'\cdot \b v'_+)\,.
\eeq
Using \eq{c31} and \eq{c33} we find 
\begin{align}
1-\unit k'\cdot \b v'_+= \frac{1-\unit k\cdot \b v_+}{1+V\cos\theta_+}=& 
\frac{1-\cos\theta\cos\theta_+-\sin\theta\sin\theta_+\cos(\phi-\phi_+)}{1-\cos\theta\cos\theta_+}\label{c36a}\\
=&\frac{1-\cos\alpha \cos\alpha_+- \sin\alpha\sin\alpha_+\cos(\beta-\beta_+)}{1-\sin\alpha\sin\beta\sin\alpha_+\sin\beta_+}\,.
\label{c36}
\end{align}
Finally, employing transformation of the time interval and the solid angle 
\beql{c37}
dt'= \frac{1}{\gamma_V}dt\,,\quad d\Omega'_+= \frac{1}{\gamma_V^2(1+V\cos\theta_+)}d\Omega_+
\eeq
we obtain that in $K'$ frame
\beql{c39}
\frac{dN}{dtd\Omega_+d\e_+}= \frac{1}{\gamma_V^2(1+V\cos\theta_+)}\frac{dN}{dt'd\Omega'_+ d\e'_+}\,.
\eeq
where $\e'$, $\e'_+$ and $\omega'$ on the right-hand-side should be replaced with the corresponding expressions in $K$ frame using \eq{c29},\eq{c31},\eq{c33},\eq{c36}.

Experimental data on dilepton production is usually represented in terms of rapidity $y_+$ and transverse momentum $p_{+\bot}$ -- which are convenient if there is no magnetic field -- in place of  energy $\e_+$ and polar angle $\alpha_+$. In the ultra-relativistic limit  
\beql{c40}
p_{+\bot}=\sqrt{ p_{+x}^2+p_{+y}^2} =\e_+\sin\alpha_+\,,\quad y_+= -\ln\tan\frac{\alpha_+}{2}\,.
\eeq 
Inverting  equations \eq{c40} yields
\beql{c41}
\e_+= p_{+\bot} \cosh y_+\,,\quad \,\sin\alpha_+= \frac{1}{\cosh y_+}\,.
\eeq
Using \eq{c41} in \eq{c29},\eq{c31},\eq{c33},\eq{c36} we can transform parameters appearing in \eq{c25} to $K$-frame as follows
\begin{align}
&\cos\theta_+=  \frac{\sin\beta_+}{\cosh y_+}\,, \qquad V= - \frac{\sin\beta}{\cosh y}\,,\qquad
\gamma_V= \frac{\cosh y}{\sqrt{\cosh^2y-\sin^2\beta}}\,,
\label{c42}\\
& \e'_+= \frac{p_{+\bot}}{\sqrt{\cosh^2y-\sin^2\beta}}(\cosh y\cosh y_+-\sin\beta\sin\beta_+)\,,  \label{c43}\\
& \omega' = \frac{\omega\sqrt{\cosh^2y-\sin^2\beta}}{\cosh y}\,, \label{c44}\\
&1-\unit k'\cdot \b v'_+= \frac{\cosh (y- y_+)- \cos(\beta-\beta_+)}{\cosh y\cosh y_+-\sin\beta\sin\beta_+}\,.\label{c45}
\end{align}
In terms of lepton's transverse momentum with respect to the collision axis $p_{+\bot}$, its rapidity $y_+$ and azimuthal angle $\beta_+$, the dilepton spectrum reads
\beql{c47}
\frac{dN_{\gamma\to \ell^+\ell^-}}{dtd^2p_{+\bot} dy_+}= \frac{dN_{\gamma\to \ell^+\ell^-}}{dtd\Omega_+d\e_+}\frac{1}{\e_+}=
\frac{\e_+'}{\gamma_V}\frac{dN_{\gamma\to \ell^+\ell^-}}{dt' d^3p'_+}\,,
 \eeq
where \eq{c39},\eq{c33} where used. The rate on the r.h.s.\ of \eq{c47} is a function of $\e'_+$, $\omega'$ etc.\ that  should be expressed through $p_{+\bot}$, $\beta_+$ and $y_+$ using \eq{c42}--\eq{c45}.

Integration over the lepton direction can be performed if we recall that in the $K'$-frame the main contribution stems from the collinear configuration $\unit k' \cdot \b v'_+\approx 1$.  Because $V\cos\theta_+\le 1$ and $1-\unit k'\cdot \b v'_+\ll 1$, it follows from \eq{c36a} that $1-\unit k\cdot \b v_+\ll 1$. Therefore, integral over the lepton directions in $K$ is still dominated by the collinear configuration $\theta_+\approx \theta$ and $\phi_+\approx \phi$. Using \eq{c33},\eq{c29} we get $\e'_+= \gamma_V\e_+(1-\cos^2\theta)= \e_+\sin\theta$ in agreement with \eq{c31}. To the same approximation $dt'\,d\e'_+$ is boost-invariant and hence we get in $K$ a formula similar to \eq{c26}
\beql{c55}
\frac{dN_{\gamma\to \ell^+\ell^-}}{dtd\e_+}= \frac{\alpha m_\ell^2}{\omega^2}\left\{ \int_x^\infty \text{Ai}(\xi)d\xi + \left( \frac{2}{x}-\varkappa x^{1/2}\right)\text{Ai}'(x)\right\}\,,
\eeq
where now
\beql{c59}
x= \left( \frac{m_\ell^3\omega}{eB\e_+\e_-\sin\theta}\right)^{2/3}\,,\qquad \varkappa = \frac{eB\omega}{m_\ell^3}\sin\theta\,.
\eeq

To calculate the dilepton spectrum produced by quarks, \eq{c47},\eq{c55} must be integrated with the equivalent photon flux $n(\omega)d\omega$ given by \eq{c3}. It is helpful to note that the equivalent photon spectrum $n(\omega) d\omega$ \eq{c3} is boost-invariant in our approximation. The differential rate per unit $\e_+$, $y_+$ and $\beta_+$ is rather bulky and we will not write it down explicitly.  The rate of  lepton production  with energy $\e_+$  reads
\begin{align}\label{c63}
\frac{dN_{q\to \ell^+\ell^-}}{dt\,d\e_+}=\frac{2z_q^2\alpha^2m_\ell^2}{\pi} \int_{\e_+}^{\omega_m}  \frac{d\omega}{\omega^3}\ln \frac{\e}{\omega[1+(\chi\e/\omega)^{1/3}]}\, \left\{ \int_x^\infty \text{Ai}(\xi)d\xi + \left( \frac{2}{x}-\varkappa x^{1/2}\right)\text{Ai}'(x)\right\} \,,
\end{align}
where, the invariant parameter $\chi$ defined in \eq{c5}  is
\beql{c67}
\chi= \frac{z_qeB\e \sin\theta}{m_q^3}\,
\eeq
and $\omega_m$ is a solution of \eq{c9}.
Integration over $\e_+$ as in \eq{c26.3} yields the total lepton rate
\beql{c69}
\frac{dN_{q\to \ell^+\ell^-}}{dt}=-\frac{2\alpha^2\, eB z_q^2}{\pi m_\ell }
\int_{2m_\ell}^{\omega_m}  \frac{d\omega}{\varkappa\omega}\ln \frac{\e}{\omega[1+(\chi\e/\omega)^{1/3}]}
\int_{(4/\varkappa )^{2/3}}^\infty\frac{2(x^{3/2}+1/\varkappa )\,\text{Ai}'(x)}{x^{11/4}(x^{3/2}-4/\varkappa )^{1/2}}dx\,.
\eeq

\section{Initial quark distribution}\label{sec:aa}

To calculate the magnetic contribution to the dilepton spectrum produced in a collision of two heavy ions with atomic weights $A_1$ and $A_2$ 
we need to convolute the initial quark distribution with the dilepton spectrum in \eq{c63} or \eq{c69}. For the differential rate per lepton's phase space $d\Gamma_+$ we have
\beql{aa-3}
\frac{dN_{A_1A_2\to \ell^+\ell^-}}{dtd\Gamma_+}= \sum_q\int \frac{dN_{A_1A_2\to qX}}{dyd^2 p_\bot}
 \frac{dN_{q\to \ell^+\ell^-}}{dtd\Gamma_+}\, d^2p_\bot dy\,.
\eeq
Sum runs over all light quarks and antiquarks. In this section we focus on the initial quark distribution.
 
 First, consider dependance of the initial quark spectrum on transverse momentum $p_\bot$. The key parameters here are two saturation momenta $Q_{s1}$ and $Q_{s2}$ of the two colliding nuclei. They depend only on quark's rapidity $y$ and total collision energy.
 At small quark's transverse momentum $p_\bot\ll \min\{Q_{s1},Q_{s2}\}$, the spectrum increases as $\frac{dN_{A_1A_2\to qX}}{dy dp_\bot}\propto p_\bot$, while at $p_\bot\gg \max\{Q_{s1},Q_{s2}\}$ it falls off as $\frac{dN_{A_1A_2\to qX}}{dydp_\bot}\propto 1/p_\bot^3$. The maximum is at $p_\bot\approx \min\{Q_{s1},Q_{s2}\}$. It is a reasonable approximation to write
\beql{aa-1}
 \frac{dN_{A_1A_2\to qX}}{dy d^2 p_\bot} \approx \delta(p_\bot-\min\{Q_{s1},Q_{s2}\})\frac{1}{2\pi p_\bot}\frac{dN_{A_1A_2\to qX}}{dy}\,.
\eeq
Using this equation in \eq{aa-3} we obtain
\begin{align}
\frac{dN_{A_1A_2\to \ell^+\ell^-}}{dtd\Gamma_+}&= \sum_q\frac{1}{2\pi}\int_0^{2\pi} d\beta \int_{-Y/2}^{Y/2} dy\,\frac{dN_{A_1A_2\to qX}}{dy }
 \frac{dN_{q\to \ell^+\ell^-}}{dtd\Gamma_+}\bigg|_{p_\bot=\min\{Q_{s1},Q_{s2}\}} \label{aa-0}\\
 &= \sum_q\frac{1}{\pi}\int_0^{2\pi} d\beta \int_{0}^{Y/2} dy\,\frac{dN_{A_1A_2\to qX}}{dy }
 \frac{dN_{q\to \ell^+\ell^-}}{dt'd\Gamma_+}\bigg|_{p_\bot=Q_{s2}}\,, \label{aa-00}
\end{align}
where we assumed for definitiveness that $Q_{s1}>Q_{s2}$ when $y>0$. $Y$ is the rapidity interval between the two ions.   

Initial differential cross section for quark  production in heavy-ion collisions in the chiral limit reads \cite{Kharzeev:2008cv,Tuchin:2012cd,Tuchin:2004rb}
\begin{align}
\frac{d\sigma_{A_1A_2\to q X}}{d^2\b\ell dy dz}&= \frac{4N_c }{(2\pi)^2 \pi^2}\int d^2b_1\int d^2b_2 \int d^2 r\int d^2 r' e^{-i\b\ell\cdot (\b r-\b r')}\frac{\b r\cdot \b r'}{r^2r'^2}
&\nonumber\\
 &\times \frac{1}{(\b r-\b r')^2}\left\{ \left[ S_1((1-z)(\b r-\b r'))S_1(z(\b r-\b r'))-1\right]\left[ S_2((1-z)(\b r-\b r'))S_2(z(\b r-\b r'))-1\right]\right\}&\nonumber\\
&\times \frac{1}{\b r^2}\left\{ \left[ S_1((1-z)\b r)S_1(z\b r)-1\right]\left[ S_2((1-z)\b r)S_2(z\b r)-1\right]\right\}\nonumber&\\
&\times \frac{1}{\b r'^2}\left\{ \left[ S_1((1-z)\b r')S_1(z\b r')-1\right]\left[ S_2((1-z)\b r')S_2(z\b r')-1\right]\right\}\,,&
\label{aa1}
\end{align}
where $\b\ell$ is the relative transverse momentum of the $q\bar q$ pair, 
\beql{aa3}
S_a(\b r) = \exp\left\{ -\frac{1}{8}\b r^2 Q_s^2(y_a,\b b_a)\right\}\,
\eeq
is the scattering matrix element, and $Q_s(y_a, \b b_a)$ is the saturation momentum of nucleus $a=1,2$.  $\b b_a$ is the impact parameter. 
Integrating \eq{aa1} over $\b\ell$ we obtain the delta function $(2\pi)^2\delta(\b r-\b r')$. Subsequent integration over $0\le r<\infty $ and $0\le z\le 1$  yields
\beql{aa5}
\frac{d\sigma_{A_1A_2\to q X}}{dy} = \frac{N_c}{24\pi^3}\int d^2b_1\int d^2b_2\left [Q_{s1}^2\ln \left( 1+\frac{Q_{s2}^2}{Q_{s1}^2}\right)+ Q_{s2}^2\ln \left( 1+\frac{Q_{s1}^2}{Q_{s2}^2}\right)\right]\,.
\eeq

It is phenomenologically reasonable approximation to treat the impact parameter dependence of $Q_s$ as a step function. Denoting by $\mathcal{S}$ the overlap area of the two ions, we obtain for the quark yield 
\beql{aa7}
\frac{dN_{A_1A_2\to q X}}{dy} = \frac{N_c \,\mathcal{S}}{24\pi^3}\left [Q_{s1}^2\ln \left( 1+\frac{Q_{s2}^2}{Q_{s1}^2}\right)+ Q_{s2}^2\ln \left( 1+\frac{Q_{s1}^2}{Q_{s2}^2}\right)\right]\,,
\eeq
where now $Q_{s}$'s are taken at $\b b_a=0$. Rapidity dependence of the saturation momentum in the center-of-mass frame is $Q_{s1,2}(y)= Q_{s0}e^{\pm\lambda y/2}$, where $\lambda$ is known phenomenological parameter \cite{GolecBiernat:1998js} and $Q_{s0}$ depends only upon the collision energy \cite{Levin:1999mw}.  With this notation we cast \eq{aa7} in form 
\beql{aa9}
\frac{dN_{A_1A_2\to q X}}{dy} = \frac{N_c \,\mathcal{S}Q_{s0}^2}{24\pi^3}\left [e^{\lambda y}\ln \left( 1+e^{-2\lambda y}\right)+ e^{-\lambda y}\ln \left( 1+e^{2\lambda y}\right)\right]\,.
\eeq

Eq.~\eq{aa9} is valid for quark rapidities not too close to the kinematic boundary  at $y= \pm Y/2$ (fragmentation regions of the nuclei). Behavior near the kinematic boundary can be inferred from the dependence of the valence quark distribution function $q_V$: when Bjorken's $x$ is close to one  $q_V(x)\propto (1-x)^3$. Using $x= (p_\bot/\sqrt{s})e^{|y|}$ with $Y= \ln(s/\mu^2)$, where $s$ is the center-of-mass energy squared and $\mu\approx 1$~GeV we  obtain 
\beql{aa11}
\frac{dN_{A_1A_2\to q X}}{dy} = \frac{N_c \,\mathcal{S}Q_{s0}^2}{24\pi^3}\,f(y)\,,
\eeq
where we introduced a shorthand notation 
\beql{aa13}
f(y)= \left [e^{\lambda y}\ln \left( 1+e^{-2\lambda y}\right)+ e^{-\lambda y}\ln \left( 1+e^{2\lambda y}\right)\right]
\left( 1-  \frac{Q_{s0}}{\mu}e^{-Y/2+|y|(1+\lambda/2) }   \right)^3\,.
\eeq

\section{Electron energy spectrum}\label{sec:xx}

To illustrate the derived results, we numerically compute the magnetic contribution to the electron energy spectrum produced in heavy ion collisions at RHIC at the central rapidity $y_+=0$. Magnetic field strength in this case is such that  $eB\gg m_e^2$. Taking also into account that  $\omega\gg m_e$ we infer from  \eq{c59} that $\varkappa \gg 1$. (This is not the case only at very small angles $\theta$ that are beyond the experimental resolution). At large $\varkappa$, the first term in the curly brackets of \eq{c55} and \eq{c63} is negligible compared to the second one.  Bearing this in mind and substituting \eq{c63} and \eq{aa9} into \eq{aa-00} we derive 
\begin{align}
 \frac{dN_{A_1A_2\to \ell^+\ell^-}}{dtd\e_+}=& \frac{N_c \,\mathcal{S}Q_{s0}^2}{3\pi^5}\alpha^2 m_\ell^2 \sum_qz_q^2\int_0^{\pi/2} d\beta \int_{0}^{Y/2} dy\,f(y)
 \nonumber\\
&\qquad\times 
 \int_{\e_+}^{\omega_m}  \frac{d\omega}{\omega^3}\ln \frac{\e}{\omega[1+(\chi\e/\omega)^{1/3}]}  \left( \frac{2}{x}-\varkappa x^{1/2}\right)\text{Ai}'(x)\,,\label{xx11}
\end{align}
where $x$, $\varkappa$ and $\chi$ are given by \eq{c59},\eq{c67} with the following substitutions:
\beql{xx13}
\e= Q_{s0}\cosh y\, e^{\lambda y/2}\,,\qquad \sin\theta=\sqrt{1-(\sin\beta/\cosh y)^2}\,.
\eeq
Our calculation is valid when $\e_+\ll \e$, i.e. in the kinematic region $\e_+\ll Q_{s0}\sim 1-1.5$~GeV. For electrons produced at $y_+=0$ at RHIC this translates into a condition $p_{+\bot}=\e_+\lesssim 0.5$~GeV. In fact, as we argue in the next section, at larger $\e_+$ 
magnetic contribution to the lepton spectrum is negligible. 

Total lepton production rate also simplifies  at large $\varkappa$. Taking integral over $x$  in \eq{c26.3} yields  \cite{Berestetsky:1982aq}
\beql{xx15}
\int_{(4/\varkappa )^{2/3}}^\infty\frac{2(x^{3/2}+1/\varkappa )\,\text{Ai}'(x)}{x^{11/4}(x^{3/2}-4/\varkappa )^{1/2}}dx= -0.38 \varkappa^{2/3}\,,\quad \varkappa\gg 1\,.
\eeq
Thus, in place of \eq{c69} we have
\beql{xx17}
\frac{dN_{q\to \ell^+\ell^-}}{dt}=0.38\frac{2\alpha^2\, eB z_q^2}{\pi m_\ell }
\int_{2m_\ell}^{\omega_m}  \frac{d\omega}{\varkappa^{1/3}\omega}\ln \frac{\e}{\omega[1+(\chi\e/\omega)^{1/3}]}\,.
\eeq
Convoluting \eq{xx17} with the hard quark spectrum \eq{aa11} yields the total electron production rate as a function of magnetic field $B$: 
\begin{align}
 \frac{dN_{A_1A_2\to e^+e^-}}{dt}=& 0.38\frac{N_c \,\mathcal{S}Q_{s0}^2}{3\pi^5}\frac{\alpha^2 eB }{ m_e} \sum_qz_q^2\int_0^{\pi/2} d\beta\int_{0}^{Y/2} dy\,f(y)
 \nonumber\\
&\qquad\times 
\int_{2m_\ell}^{\omega_m}  \frac{d\omega}{\varkappa^{1/3}\omega}\ln \frac{\e}{\omega[1+(\chi\e/\omega)^{1/3}]}\,.\label{xx19}
\end{align}

\begin{figure}[ht]
      \includegraphics[height=5cm]{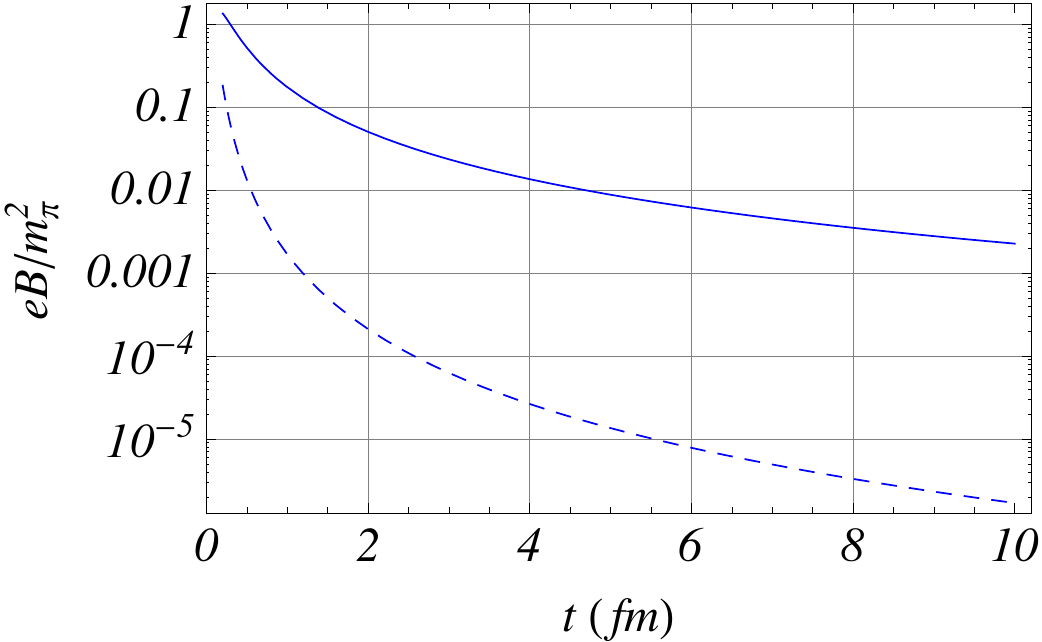} 
  \caption{Time-dependence of magnetic field at midrapidity after a collision of two Gold nuclei at $\sqrt{s_{NN}}=200$~GeV. Solid line: \eq{xx39} with $\sigma=5.8$~MeV\cite{Ding:2010ga}, dashed line: \eq{xx41}. }
\label{fig:xx3}
\end{figure}

All equations that we derived so far pertain to lepton production in static magnetic field.
 In fact, magnetic field does change with time albeit adiabatically. Its time-dependence at the central rapidity $y=0$ is approximately given by \cite{Tuchin:2010vs,Tuchin:2013ie}
\beql{xx39}
 e\b B(t)=  \unit y\,\frac{\alpha ZR_A\sigma}{t^2}\exp\left\{-\frac{R_A^2\sigma}{4t}\right\}\,,
 \eeq
where $\sigma$ is QGP electrical conductivity, $Z$ and $R_A$ are nuclear charge and radius correspondingly. Derivation of \eq{xx39}  assumes that QGP conductivity is constant. In expanding medium $\sigma$ is a function of time, e.g.\ in Bjorken scenario \cite{Bjorken:1982qr} $\sigma\sim t^{-1/3}$. However it has only mild effect on the time-dependence of magnetic field given by \eq{xx39}.  Taking for Gold nucleus $Z=79$, $R_A=6.5$~fm we obtain that at $t=0.2$~fm (which is $\sim 1/Q_s$) $eB=1.3~m_\pi^2$ in agreement with earlier estimates \cite{Kharzeev:2007jp}. However, time-dependence of \eq{xx39} is significantly different from the one found in \cite{Kharzeev:2007jp} because it takes into account the electromagnetic response of QGP.  The time-dependence of magnetic field in conducting medium is shown by a solid line in \fig{fig:xx3}.
It is seen that even at $t=10$~fm magnetic field is about two orders of magnitude larger than the Schwinger field for electrons: $eB_c/m_\pi^2= (m_e/m_\pi)^2= 1.3\cdot 10^{-5}$. For comparison, we also calculated the dilepton spectrum without the medium effect on magnetic field. In this case, time-dependence of magnetic field can be modeled by boosted Coulomb field as
\beql{xx41}
e\b B = \unit y\, \frac{2\alpha Z\,R_A\gamma }{(R_A^2+\gamma^2 t^2)^{3/2}}\,,
\eeq
where $\gamma=\sqrt{s_{NN}}/2m_N$ is the Lorentz factor. At RHIC heavy-ions are collided at 200~GeV per nucleon, hence $\gamma=100$. The time-dependence of magnetic field in vacuum is shown by a dashed line in \fig{fig:xx3}.

\begin{figure}[ht]
         \includegraphics[height=6cm]{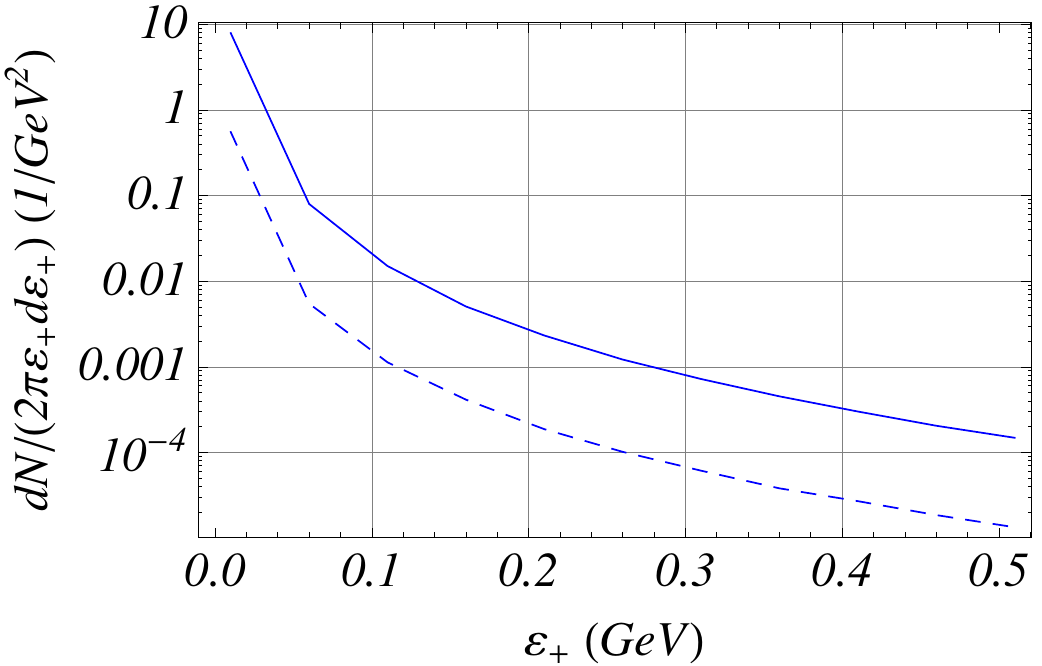} 
 \caption{Spectrum of electrons produced in magnetic field by hard quarks during the QGP life-time. It coincides with the distribution in $p_{+\bot}$  at midrapidity $y_+=0$. Time dependence of magnetic field is given by \eq{xx39} (solid line) and by \eq{xx41} (dashed line).}
\label{fig:xx5}
\end{figure}

To calculate the total time-integrated yield of electrons substitute \eq{xx39} (in medium)  or \eq{xx41} (in vacuum) into \eq{xx11} and \eq{xx19} and integrate over time in the interval $0.2\le t(\text{fm})\le 10$. The result of the calculation is displayed in \fig{fig:xx5}.
Integrating over the electron energy $\e_+$ we obtain that the total number of leptons produced per event $N_e=0.049$ (in medium), and $N_e=0.0033$ (in vacuum). These numbers as well as \fig{fig:xx5} represent a magnetic contribution to $e^+e^-$ yield at RHIC \emph{only} due to the square of the amplitude of  \fig{fig:int1}. It does not take into account other contributions represented by diagrams in \fig{fig:int2}.

%
%
\section{Comparison of  magnetic and conventional photon decay  mechanisms}\label{sec:n}

\begin{figure}[ht]
      \includegraphics[height=5cm]{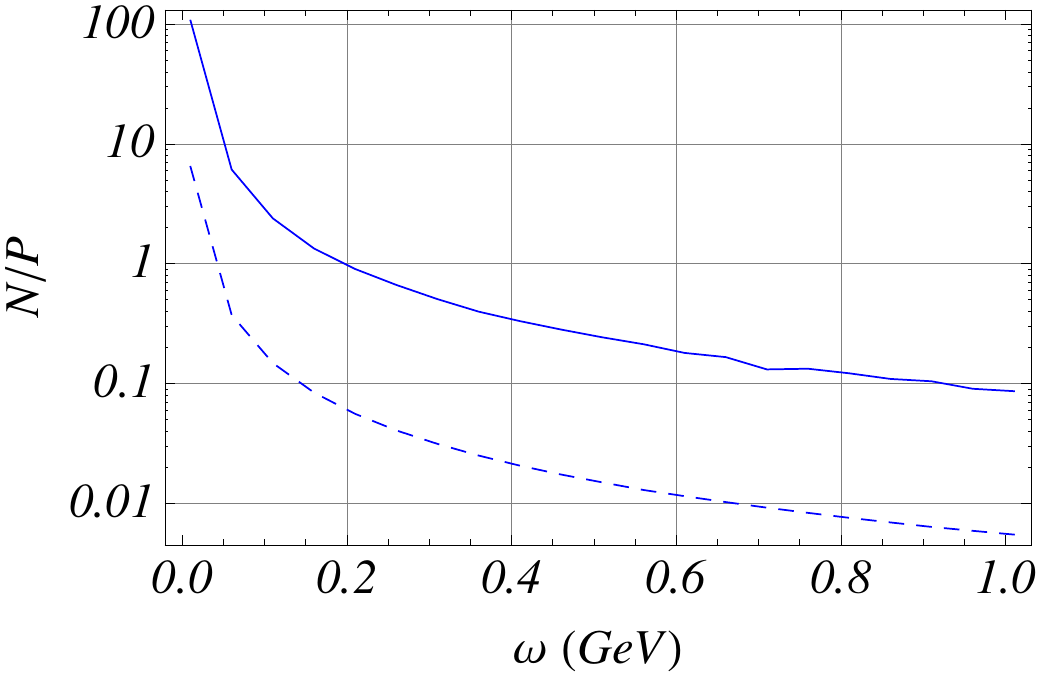} 
  \caption{Ratio of dilepton production yields:  via photon splitting in magnetic field $N_{\gamma\to \ell^+\ell^-}$ and via decay of virtual photon  without magnetic field $P_{q\to \ell^+\ell^-}$. Solid and dashed lines correspond to the time-dependence of magnetic field according to \eq{xx39} and \eq{xx41}. Photon's rapidity $y=0$. 
  }
\label{fig:n1}
\end{figure}

In the absence of magnetic field, dilepton production by a fast quark is related to the dilepton production by a virtual photon as follows  
\beql{n1}
\frac{dN_{q\to \ell^+\ell^-}}{d\Gamma_+} = \frac{2\alpha}{3\pi }\int \frac{dM }{M} \frac{dN_{\gamma^*}}{d\Gamma_+}\,,
\eeq
provided that the invariant mass $M$ satisfies $M\gg m_\ell$.  If $\omega\gg M$ then the photon spectrum is $M$-independent and can be approximated by that of real photons. In that case we obtain 
\beql{n3}
\frac{dN_{q\to \ell^+\ell^-}}{d\Gamma_+} = \frac{2\alpha}{3\pi }\ln \frac{\omega}{m_\ell}\,\frac{dN_{\gamma}}{d\Gamma_+} = P_{\gamma^*\to \ell^+\ell^-} \frac{dN_{\gamma}}{d\Gamma_+} \,,
\eeq
where $P_{q\to \ell^+\ell^-}$ describes the probability  to produce dilepton via  photon decay. This quantity should be compared with $N_{\gamma\to \ell^+\ell^-}$, which describes such a probability in magnetic field. If one tries to  reconstruct the photon spectrum
using \eq{n1}, then the result will be incorrect as it misses an important $B$-dependent contribution.\footnote{However, if the photon decay vertex is measured, it implies that photon decay happened long after the disappearance of QGP and the magnetic field it supported. Such process can certainly be described by \eq{n1}.} To demonstrate how different are these contributions, we plotted their ratio in \fig{fig:n1} for  azimuthal angle $\beta=0$, i.e.\  perpendicular to $\b B$. (We neglect small variations with $\beta$).
Magnetic  contribution dominates at low frequencies. At midrapidity $\omega = k_\bot$ and we can infer from \fig{fig:n1} that at  $k_\bot=\omega<0.2$~GeV the contribution  to the dilepton yield induced by magnetic field dominates over the conventional virtual photon splitting. Even at $k_\bot \sim 0.6$~GeV it contributes as much as  20\% to the yield.  
Dilepton production in magnetic field certainly contributes to resolution of the puzzle of  enhancement of dilepton  production reported in \cite{Adare:2009qk}. However, as has been already mentioned, the exact value of magnetic field contribution is hard to pin down without accurate knowledge of the magnetic field dynamics.

\section{Summary}\label{sec:sum}

It is hardly surprising that magnetic field generated in heavy ion collisions has a profound impact on dilepton production. Indeed, magnetic field strength by far exceeds the critical Schwinger's value during the entire QGP lifetime. In the present work we calculated a contribution to  dilepton spectrum in the region $T<\e_+<Q_s$ due to magnetic field as a convolution of three factors: the initial hard (i.e.\ non-thermal) quark distribution, equivalent photon flux and photon decay rate. The last two factors exhibit very strong dependence on magnetic field. Because momentum is conserved only in the direction of  magnetic field, the notion of invariant mass applies only to the part of the spectrum independent of magnetic field. Consequently, we plotted the electron spectra in  \fig{fig:xx3} as a function of electron's energy. We also derived formulas for  fully differential distribution of leptons.  We argued that the magnetic contribution is important at electron energies below $\sim 0.5$~GeV at midrapidity at RHIC. In fact it  becomes the dominant source of dileptons at lower electron energies, see \fig{fig:n1}.

In the region $\e_+<T$ contribution of soft quarks, i.e.\ those quarks that are part of the QGP cannot be neglected and must be added to the magnetic contribution of hard  quarks  computed in this paper. Its calculation however is much more complicated because the equivalent photon approximation is no longer applicable. We are planning to discuss this contribution elsewhere.

\acknowledgments
I  am grateful to Yoshimasa Hidaka and Kazunori Itakura for interesting discussions that initiated this work and  to Thomas Hemmick for useful correspondence. This work  was supported in part by the U.S. Department of Energy under Grant No.\ DE-FG02-87ER40371.



\begin{thebibliography}{80}

\bibitem{Itakura:2013cia} 
  K.~Itakura,
  J.\ Phys.\ Conf.\ Ser.\  {\bf 422}, 012029 (2013).

\bibitem{Tuchin:2012mf} 
  K.~Tuchin,
  Phys.\ Rev.\ C {\bf 87}, 024912 (2013)
  [arXiv:1206.0485 [hep-ph]].

\bibitem{Tuchin:2010gx}
  K.~Tuchin,
  Phys.\ Rev.\  {\bf C83}, 017901 (2011).
  [arXiv:1008.1604 [nucl-th]].

\bibitem{Yee:2013qma} 
  H.~-U.~Yee,
  arXiv:1303.3571 [nucl-th].

\bibitem{Adare:2008ab} 
  A.~Adare {\it et al.}  [PHENIX Collaboration],
  Phys.\ Rev.\ Lett.\  {\bf 104}, 132301 (2010).
  
  
\bibitem{Adare:2009qk} 
  A.~Adare {\it et al.}  [PHENIX Collaboration],
  Phys.\ Rev.\ C {\bf 81}, 034911 (2010)
  [arXiv:0912.0244 [nucl-ex]].
  
\bibitem{Kolb:2003dz} 
  P.~F.~Kolb and U.~W.~Heinz,
  In *Hwa, R.C. (ed.) et al.: Quark gluon plasma* 634-714
  [nucl-th/0305084].
  
\bibitem{Kharzeev:2008cv} 
  D.~Kharzeev, E.~Levin, M.~Nardi and K.~Tuchin,
  Nucl.\ Phys.\ A {\bf 826}, 230 (2009)
  [arXiv:0809.2933 [hep-ph]].
  
\bibitem{Kovchegov:2006qn}
  Y.~V.~Kovchegov and K.~Tuchin,
  Phys.\ Rev.\ D {\bf 74}, 054014 (2006)
  [arXiv:hep-ph/0603055].


\bibitem{Gelis:2003vh}
  F.~Gelis and R.~Venugopalan,
  Phys.\ Rev.\ D {\bf 69}, 014019 (2004)
  [arXiv:hep-ph/0310090].

\bibitem{Blaizot:2004wv}
  J.~P.~Blaizot, F.~Gelis and R.~Venugopalan,
  Nucl.\ Phys.\ A {\bf 743}, 57 (2004)
  [arXiv:hep-ph/0402257].

\bibitem{Kopeliovich:2002yv} 
  B.~Z.~Kopeliovich and A.~V.~Tarasov,
  Nucl.\ Phys.\ A {\bf 710}, 180 (2002)
  [hep-ph/0205151].

\bibitem{Kharzeev:2003sk}
  D.~Kharzeev and K.~Tuchin,
  Nucl.\ Phys.\  A {\bf 735}, 248 (2004)
  [arXiv:hep-ph/0310358].

\bibitem{Tuchin:2004rb} 
  K.~Tuchin,
  Phys.\ Lett.\ B {\bf 593}, 66 (2004)
  [hep-ph/0401022].
 
  
\bibitem{Baier:1973} 
  V.~N.~Baier and V.~M.~Katkov,
  Sov.\ Phys.\ Dokl.\  {\bf 17}, 1068 (1973).

\bibitem{Berestetsky:1982aq}
  V.~B.~Berestetsky, E.~M.~Lifshitz and L.~P.~Pitaevsky,
 ``Quantum Electrodynamics,'' \S90, 
{\it  Oxford, Uk: Pergamon (1982) 652 P.\ (Course Of Theoretical Physics, 4)}.

  
\bibitem{Tuchin:2012cd} 
  K.~Tuchin,
  Nucl.\ Phys.\ A {\bf 899}, 44 (2013)
  [arXiv:1209.0799 [hep-ph]].
  
\bibitem{GolecBiernat:1998js} 
  K.~J.~Golec-Biernat and M.~Wusthoff,
  Phys.\ Rev.\ D {\bf 59}, 014017 (1998)
  [hep-ph/9807513].

\bibitem{Levin:1999mw} 
  E.~Levin and K.~Tuchin,
  Nucl.\ Phys.\ B {\bf 573}, 833 (2000)
  [hep-ph/9908317].

  

  
\bibitem{Tuchin:2013ie} 
  K.~Tuchin, Adv.\ High Energy Phys.\ (in press), 
  [arXiv:1301.0099 [hep-ph]].
  
\bibitem{Tuchin:2010vs}
  K.~Tuchin,
  Phys.\ Rev.\  {\bf C82}, 034904 (2010).
  [arXiv:1006.3051 [nucl-th]].
  
   
\bibitem{Bjorken:1982qr} 
  J.~D.~Bjorken,
  Phys.\ Rev.\ D {\bf 27}, 140 (1983).
  
\bibitem{Kharzeev:2007jp}
  D.~E.~Kharzeev, L.~D.~McLerran and H.~J.~Warringa,
  Nucl.\ Phys.\  A {\bf 803}, 227 (2008).
  
\bibitem{Ding:2010ga} 
  H.~-T.~Ding, A.~Francis, O.~Kaczmarek, F.~Karsch, E.~Laermann and W.~Soeldner,
  Phys.\ Rev.\ D {\bf 83}, 034504 (2011)
  [arXiv:1012.4963 [hep-lat]].

   

\end{thebibliography}
\end{document}